\documentclass[twoside]{article}
\usepackage[accepted]{aistats2e}
\usepackage{amsmath,amssymb}
\usepackage[round]{natbib}
\usepackage{url}
\input{Definitions}

\newcommand{\email}{\begingroup\urlstyle{sf}\small\Url}
\newcommand{\x}{$\!\!\!\!$}

\title{Variable Metric Stochastic Approximation Theory}

\author{
{\bf Peter Sunehag\thanks{~Also an adjunct at the second author's affiliation.}}\\
\email{Peter.Sunehag@nicta.com.au}\\
NICTA, Locked Bag 8001\\
Canberra ACT 2601\\
Australia\\
\And
{\bf Jochen Trumpf}\\
\email{Jochen.Trumpf@anu.edu.au}\\
Dept.\ of Information Engr.\\
Australian National Univ.\\
Canberra, 
Australia$\!\!\!$\\
\And
\x {\bf S.V$\!$.\,N. Vishwanathan}\\
\x \email{vishy@stat.purdue.edu}\\
\x Statistics Department\\
\x Purdue University\\ 
\x West Lafayette, IN\\ 
\And
{\bf Nicol N. Schraudolph}~~~\\
{\email{schraudolph@adaptivetools.com}}~~~\\
adaptive tools AG~~~\\
Canberra ACT 2602~~~\\
Australia~~~
}

\begin{document}
\runningtitle{Variable Metric Stochastic Approximation Theory}%
\runningauthor{Sunehag, Trumpf, Vishwanathan, and Schraudolph}%
\checkheadings
\maketitle

\begin{abstract}
  We provide a variable metric stochastic approximation theory. In doing so, we provide a convergence theory
  for a large class of online variable metric methods including the recently introduced online versions of the BFGS algorithm and
  its limited-memory LBFGS variant. We also discuss the implications of our results for learning from
  expert advice.
    \end{abstract}

\section{Introduction}

We begin by introducing online optimization and in particular stochastic gradient descent methods.

\subsection{Online Optimization}
There exists an abundance of applications that can lead us to online
optimization problems where we are trying to find the minimum of a
data-dependent function $C: \mathbb{R}^n \to \mathbb{R}, w \mapsto C(w)
= C(w,Z)$, where $Z$ represents data or a probability distribution that
generates data. During an online optimization procedure, a sequence of parameters $w_t$,
$t = 1, 2, \ldots$ is created by using an update rule for how to define
$w_{t+1}$ given the earlier parameters and the new data, $z_t$, that has
arrived at time $t$.
Given a sequence of data samples $z_1,...,z_t$ drawn from a fixed distribution, we will
use the notation $C_t(\cdot)=\frac{1}{t}\sum_{i=1}^t C(\cdot,z_i)$ for the empirical objective which is
the average of online/stochastic/instantaneous objectives $C(\cdot,z)$.
We will refer to $C(\cdot)=\EE_z C(\cdot,z)$ as the true objective.

\subsection{Stochastic Gradient Descent}
If one has defined a metric $\kappa$ on the parameter space by supplying a dot product, and if
the online objectives are differentiable, then we can use the gradients $\nabla^\kappa$ with respect to that metric
to define the update equation
\begin{equation}
w_{t+1}=w_t-a_t\nabla^\kappa_w C(w_t,z_t), a_t>0.
\end{equation}
If one uses metrics defined by different dot products for different $t$, then one can let $\nabla$ denote the gradient with respect to the standard Euclidean dot product  and instead let the updates take the form
\begin{equation}\label{eq:theupdate}
w_{t+1}=w_t-a_tB_t\nabla_w C(w_t,z_t),
\end{equation}
where the $B_t$ are positive definite and symmetric matrices.
One example of considering variable metrics is the study of information geometry
\citep{AmaNag93}, where the Fisher information matrix is used to define a metric tensor on
a family of probability distributions. More specific examples will be provided in Section \ref{sec:Objectives} below.

\subsection{Outline and Summary}
\label{sec:Outline}

We investigate the theoretical foundations for using
online updates that include scaling matrices, that is stochastic gradient descent where the gradients are taken with respect to time-varying metrics. Among other results, this provides a convergence
proof for a large class of variable metric methods including the recent online (L)BFGS algorithm \citep{SchYuGue07}.
In Section \ref{sec:sat} we employ the
Robbins-Siegmund theorem to prove $O(1/t)$ convergence in
function values $C(w_t)$ for a class of functions that is useful in machine
learning. This is the best
possible rate \citep{BotLec05,Amari98, Mur98}, limited only by the rate at which information arrives.
Under weaker assumptions we show almost sure convergence
without rate for a larger class. Our results extend those in \citet{BotLec05}
by not demanding that the metrics converge to an asymptotic metric.

We first introduce motivating application areas in Section \ref{sec:Objectives}. Then in Section \ref{sec:sat} we
both provide a background in stochastic approximation theory and our new theorems.
 We apply our
theorems to prove convergence for online BFGS in Section \ref{sec:bfgs}.
Furthermore, we consider implications in the area of learning from expert advice in Section \ref{sec:expert}.
The paper concludes with a discussion in Section \ref{sec:conclusion}.

\section{Examples}
\label{sec:Objectives}
 In this section we provide different objective functions and describe the problem settings they appear in.

\subsection{Online Risk Minimization}

The goal of many machine learning algorithms is to minimize the risk
$\EE_{z} l(\inner{w}{x}, y)$, where the expectation is taken with respect to a fixed
but unknown probability distribution $Z$ that generates instance-label pairs
$z := (x, y)$. The loss $l$ is a non-negative convex function of the
parameters $w$ and measures the discrepancy between the labels
$y$ and the predictions arising from $x$ and $w$ via their inner
product $\inner{w}{x}$ (often the Euclidean dot product if
$x, w \in \mathbb{R}^k$).

In the absence of complete knowledge about the underlying distribution,
an empirical sample $Z = \{(x_{i}, y_{i}), i = 1, \ldots, n\}$\footnote{With
some abuse of notation we use $Z$ to represent either a data set or a
distribution that generates data.}
is often used to minimize the regularized empirical risk\begin{align}
  \label{eq:regrisk}
   \frac{c}{2} \|w\|^2 + \frac{1}{n} \sum_{i=1}^{n}
  l(\inner{w}{x_{i}}, y_i)
\end{align}
where the $L_2$-regularization term $\frac{c}{2}\|w\|^2$ is introduced for well-posedness.
With larger $n$ one can use smaller $c>0$. The true regularized risk that is estimated by \eqref{eq:regrisk}
is $\frac{c}{2}\|w\|^2+\EE_{z} l(\inner{w}{x}, y)$.

Batch optimization algorithms, including quasi-Newton and bundle-based
methods, are available and widely used to minimize \eqref{eq:regrisk}, but
they are computationally expensive. Gradient-based batch methods may also
fail to converge if the loss is non-smooth. Therefore, online optimization
methods that work with small subsamples of training data have received
considerable attention recently
\citep{KivWar97,Schraudolph02,AzoWar01,ShaSinSre07,SchYuGue07}.


In the online setting, we replace the objective \eqref{eq:regrisk}
with approximations based on subsets (``mini-batches'') of the samples:
\begin{align}
  \label{eq:regrisk-stoc}
   \frac{c}{2} \|w\|^2 + \frac{1}{b} \sum_{(x,y) \in Z_{t}}
  l(\inner{w}{x}, y),
\end{align}
where $Z_t \subset Z$ with $|Z_t| = b \ll n$. Furthermore, in online learning we can consider using $c=0$
and thereby aiming directly at minimizing the true objective $\EE_{z} l(\inner{w}{x}, y)$.  During online optimization, a sequence of parameters
$w_t$, $t = 1, 2, \ldots$ arises from an update rule that computes $w_{t+1}$
from the previous state and the new information at time $t$.
In addition to alleviating the high computational cost of batch
methods, the online setting also arises when the data itself is
streaming, that is, we are receiving partial information about
$C(\cdot)$ in a sequence of small packages.

Including second-order information in the online optimization procedure can accelerate the convergence
\citep{SchYuGue07}. This is particularly true in a setting where only one pass through a dataset is performed. \citet{BotLec05} and \citet{Mur98} point out that minimizing the empirical objective is different from minimizing the true objective, and show that the result of an online second-order gradient descent procedure
can be as close to the true optimum as the minimum of the empirical objective.

\subsection{Filtering}
The goal in filtering is to separate the signal from the noise in a stream of data. Kalman algorithms
in particular use the Euclidean distance (sum-squared loss) and track
the minimizer of the empirical objectives $C_t(w)=\sum_{i=1}^t (y_t-w\cdot x_t)^2$.
The inverse of the Hessian of $C_t$  is used as $B_t$ , $a_t=1$ and $w_0=0$ for the update in \eqref{eq:theupdate}. $B_{t+1}$ is found from $B_t$ with an update whose cost is order $n^2$.
The result is that $w_{t}=\argmin_w C_t(w)$.  Therefore, if we have a fixed distribution the sequence
will converge to the optimal parameters.

The same algorithm can be extended to a more general setting where the sum-squared loss is replaced by arbitrary
convex functions. The resulting algorithm was called the online Newton-step algorithm by \citet{HazAgaSat07}
and described as an approximate ``follow the leader'' algorithm, \ie an algorithm that approximately follows
the optimum that the Kalman filter tracks exactly for the sum-squared loss.

\subsection{Learning from Expert Advice with Bregman Divergences}\label{breg}
The general convex optimization framework by \citet{Zinkevich03} that \citet{HazAgaSat07} worked in is also related to the expert advice framework by \citet{AzoWar01}. In the expert advice framework one encounters
a sequence of loss functions $L_t$ ($L_t(w)=C(w,z_t)$ here) and one wants to perform the implicitly defined update
\begin{equation}
w_{t+1}=\argmin_w \Delta_H(w,w_t)+a_t L_t(w)
\end{equation}
where $\Delta_H$ is the Bregman divergence
\begin{equation}
\Delta_H(p,q)=H(p)-H(q)+\nabla H(q)\cdot(p-q)
\end{equation}
defined by the differentiable convex function $H$.
The squared EÄuclidean distance is an example of a Bregman divergence, corresponding to $H(p)=\frac{1}{2}\|p\|_2^2$. The goal for \citet{AzoWar01}, as it was for \citet{Zinkevich03}, was to have a small total accumulated loss $\sum_{j=1}^t L_j(w_j)$ and
in particular small regret, $\sum_{j=1}^t L_j(w_j)-\min_w\sum_{j=1}^t L_j(w)$.
To derive an explicit update, \citet{AzoWar01} differentiate the expression to be minimized
\begin{align}
\nabla (\Delta_H(w,w_t) & + a_t L_t(w)) = \\ & h(w)-h(w_t)+a_t\nabla L_t(w) \nonumber
\end{align}
and by
using the approximation $\nabla L_t(w)\approx\nabla L_t(w_t)$, they arrive at the updates
\begin{equation}\label{manfred}
w_{t+1}=h^{-1}(h(w_t)-a_t\nabla L_t(w_t)).
\end{equation}
This reduces to online gradient descent, $w_{t+1}=w_t-\nabla L_t(w_t)$, for the case $h(w)=w$,
\ie when $H(w)=\frac{1}{2}\|w\|_2^2$.

\citet{AzoWar01} in particular consider loss functions of a special form, namely the case where
$L_t(w)=\Delta_G(x_t\cdot w,g^{-1}(y_t))$ where $g=\nabla G$ is an increasing continuous function
from $\mathbb{R}$ to itself and where $(x_t,y_t)=z_t$ is an example.
This results in simpler updates since it implies that $\nabla L_t(w)=(g(w\cdot x_t)-y_t)x_t$.
Note that, given a parametrization of $\mathbb{R}^d$, a one-dimensional transfer function $g$ can
be used to define a $d$-dimensional continuous bijection by applying it coordinate-wise.

It is interesting to compare \eqref{manfred} to a reparametrization where one makes a coordinate
change $\gamma=h(w)$. Then \eqref{manfred} can be written as
\begin{equation}\label{manfred2}
h(w_{t+1})=h(w_t)-a_t\nabla_w L_t(w_t).
\end{equation}
The one obstacle to identifying \eqref{manfred2} with
\begin{equation}\label{reparam}
\gamma_{t+1}=\gamma_t-a_t\nabla_\gamma \tilde{L}_t(\gamma_t)
\end{equation}
where $\tilde{L}_t(\gamma)=L(h^{-1}(\gamma))$, is that the gradient of $L_t$ is taken with respect to $w$.
We know that $\nabla_\gamma \tilde{L}_t(\gamma_t)=B_t \nabla_w L_t(w_t)$, where $B_t$ is the inverse
Hessian of $H$ at $w_t$, in other words the inverse Jacobian of $h$.

A reason for doing a reparametrization is that the loss functions $L_t$ might not satisfy the conditions
that are needed for convergence of stochastic gradient descent, but it can be possible to find a transfer
function $h$ such that $\tilde{L}_t$ does. In that sense, the update \eqref{manfred} can be a way of trying to
do this even if we only have the gradients $\nabla_w L_t$ and do not want to calculate the
matrices $B_t$. Our main theorems will tell us when it is acceptable to omit $B_t$, and thereby use an SGD
update with respect to varying metrics.

\section{Stochastic Approximation Theory}
\label{sec:sat}

\citet{RobMon51} proved a theorem that implies convergence for
one-dimensional stochastic gradient descent; \citet{Blum54} generalized
it to the multivariate case. \citet{RobSie71} achieved a stronger result of
wider applicability in supermartingale theory. Here we extend the known
convergence results \citep{BotLec05} in two ways: a) We prove that
updates that include scaling matrices with eigenvalues bounded by
positive constants from above and below will converge almost surely;
b) under slightly stronger assumptions we obtain a $O(1/t)$ rate of
convergence in the function values.

\subsection{The Multivariate Robbins-Monro Procedure}

Suppose that $\nabla C = f: \mathbb{R}^k \to \mathbb{R}^k$ is an unknown
continuous function that we want to find a root of. Furthermore, suppose
that there is a unique root and denote it by $w^*$. Given an initial
estimate $w_1$, the procedure constructs a sequence of estimates $w_t$
such that $w_t\to w^*$ as $t \to \infty$.  For any random vector $X$ let
$\EE_t(X)$ be the conditional expectation given $w_1, \ldots, w_t$.
Given $w_1, \ldots, w_t$, we assume that we observe an unbiased estimate
$Y_t$ of $f(w_t)=\nabla C(w_t)$, \ie $\EE Y_t = f(w_t)$.
Given $Y_t$ and $w_t$ we define $w_{t+1}$ by
\begin{align}
w_{t+1} = w_t - a_t Y_t,
\end{align}
where $a_t > 0$ for all $t$, $\sum a_t = \infty$ and $\sum a_t^2 <
\infty$. The $Y_t$ are assumed to be drawn from a family
$Y(x)$ of random vectors defined for all $x\in\mathbb{R}^k$, and $Y_t$
is distributed as $Y(w_t)$.  To ensure that $w_t$ converges to $w^*$
almost surely it is sufficient to assume that there are finite constants
$A$ and $B$ such that $\EE\|Y(x)\|^2 \leq A + B\|x-w^*\|^2$ for all $x$,
and that for all $\varepsilon>0$
\begin{align}
\label{origcond}
\inf\{(x - w^*)^T f(x): \varepsilon<\|x - w^*\|<\varepsilon^{-1}\} > 0.
\end{align}
For instance, strictly convex functions satisfy \eqref{origcond}.
This classical convergence result is implied by the following
theorem on almost positive supermartingales, which we will also use to prove our results:
\begin{theorem}[\citealp{RobSie71}]
Let $(\Omega,\mathcal{F},P)$ be a probability space and
$\mathcal{F}_1\subseteq\mathcal{F}_2\subseteq \ldots$ a sequence of
sub-$\sigma$-fields of $\mathcal{F}$. Let $U_t,\beta_t,\xi_t$ and
$\zeta_t$, $t=1,2,\ldots$ be non-negative $\mathcal{F}_t$-measurable
random variables such that
\begin{align}
\EE(U_{t+1}\ |\ \mathcal{F}_t)\leq (1+\beta_t)U_t+\xi_t-\zeta_t,\ t = 1, 2, \ldots
\end{align}
Then on the set $\{\sum_t\beta_t<\infty, \sum_t\xi_t<\infty\}$, $U_t$ converges
almost surely to a random variable, and $\sum_t\zeta_t<\infty$ almost surely.
\label{thm:RS}
\end{theorem}

\subsection{Convergence for Updates with Scaling Matrices}
\label{sec:convg}
\citet{BotLec05} previously presented results on the rate of convergence
in parameter space (using the Euclidean norm) of online updates with
\emph{convergent} scaling matrices, and remark that bounds on the eigenvalues
of the scaling matrix will be essential to extending convergence guarantees
beyond that. Since recent online quasi-Newton methods \citep{SchYuGue07}
do not provide convergence of their scaling matrices, we now employ the
Robbins-Siegmund theorem to prove that such updates are still guaranteed
almost sure convergence, provided the spectrum of their (possibly non-convergent)
scaling matrices is uniformly bounded from above by a finite constant and from
below by a strictly positive constant.

\begin{theorem}\label{mainthm}
Let $C:\mathbb{R}^n\to\mathbb{R}$ be a twice differentiable cost function with
unique minimum $w^*$, and let
\vspace{-\baselineskip}
\begin{align}
w_{t+1}=w_t-a_tB_tY_t,
\end{align}
where $B_t$ is symmetric and only depends on information available at time $t$. \\
Then $w_t$ converges to $w^*$ almost surely if the following conditions hold:
\vspace{-1ex}
\begin{description}
\setlength{\itemsep}{0pt}
\item[C.1] $(\forall t) ~\EE_t Y_t = \nabla_w C(w_t)$;
\item[C.2] $(\exists K) \,(\forall w) ~\|\nabla_w^2 C(w)\|\leq 2K$;
\item[C.3] $(\forall \delta > 0) ~\inf_{C(w) - C(w^*) > \delta} \|\nabla_wC(w)\| > 0$;
\item[C.4] $(\exists A,B) \,(\forall t) ~\EE \|Y_t\|^2\leq A+BC(w_t)$;
\item[C.5] $(\exists~ m, M \!: 0 < m < M < \infty) \,(\forall t) ~mI \prec B_t \prec MI$,
 ~where $I$ is the identity matrix;
\item[C.6] $\sum_t a_t^2<\infty$ ~and~ $\sum_t a_t=\infty$.
\end{description}
\end{theorem}
\begin{proof}
Since $C$ is twice differentiable and has bounded Hessian (C.2) we can use Taylor
expansion and the upper eigenvalue bound (C.5) to prove that
\begin{equation}
\label{keyeq1}
C(w_{t+1}) = C(w_t - a_t B_t Y_t) \leq
\end{equation}
\begin{equation*}
 C(w_t) - a_t [\nabla_w C(w_t)]^T B_t Y_t + K M^2 a_t^2 \|Y_t\|^2
\end{equation*}
which implies, using (C.1) and (C.4), that
\begin{equation}\label{keyeq2}
  \EE_t C(w_{t+1}) \leq C(w_t) + K M^2 a_t^2 [A + B C(w_t)] -
  \end{equation}
\begin{equation*}
a_t [\nabla_w C(w_t)]^T B_t \nabla_w C(w_t).
\end{equation*}
If we let $U_t = C(w_t)$ and merge the terms containing $U_t$ it follows that $\EE_t U_{t+1} \leq$
\begin{equation}\label{keyeq}
U_t(1 + a_t^2 B K M^2) + A K M^2 a_t^2 - m a_t \|\nabla_w C(w_t)\|^2.
\end{equation}
Since $\sum_t a_t^2 < \infty$ (C.6), the Robbins-Siegmund theorem can now be applied. We find that
\begin{align}
\sum_t a_t \|\nabla_wC(w_t)\|^2 < \infty.
\end{align}
Since $\sum a_t = \infty$ (C.6) it follows from (C.3) that
\begin{align}
\|(\nabla_w C(w_t))\|^2 \to 0
\end{align}
and that $C(w_t) \to C(w^*) $ as $t \to \infty$.
\end{proof}

\begin{remark}
  The assumption that $C$ is twice differentiable is only needed for the Taylor
  expansion we use to obtain \eqref{keyeq1}.  If we have such a property from
  elsewhere we do not need twice-differentiability.
\end{remark}

\subsection{Asymptotic Rates}
\label{sec:asymp}

Consider the situation described in Theorem \ref{mainthm}. We now strengthen
assumption (C.3), which demands that the function $C$ is not so flat around the
minimum that we may never approach it, to instead assuming that
\begin{align}\label{cond}
\frac{C(w_t)-C(w^*)}{\|\nabla C(w_t)\|^2}\leq D<\infty ~~\forall t.
\end{align}
Condition \eqref{cond} is implied by strong convexity if we know that the $w_t$
tend to the optimum $w^*$.
Since $\nabla C(w^*)=0$ we can use first-order Taylor expansion of $\nabla C$
around $w^*$ to approximate $\|\nabla C(w)\|^2$ by
$(w-w^*)^T\nabla^2 C(w^*)(w-w^*)$.

We will also modify assumption (C.4) by setting $B=0$. Theorem \ref{mainthm} guarantees (under the weaker conditions) that the procedure
will almost surely generate a converging sequence which is therefore contained in some ball around $w^*$. This makes the
new condition reasonable. \cite{Bottou98} contains a more elaborate discussion on what is there called "Global Confinement".

We need a result on what the expected improvement is, given the step size $a_t$, the uniform bound on the Hessian, and the uniform eigenvalue bounds.
The key to achieving this is \eqref{keyeq1}.
This section's counterpart to \eqref{keyeq2} under the new conditions is
\begin{equation}
\label{neweq}
\EE_t C(w_{t+1}) - C(w^*) \leq
\end{equation}
\begin{equation*}
[C(w_t) - C(w^*)] (1 - a_t m/D) + A K M^2 a_t^2.
\end{equation*}
We want to know the rate of the sequence $\EE C(w_t) - \inf_{\tilde{w}} C(\tilde{w})$, and will
use the fact that  taking the unconditional expectation of \eqref{neweq} yields the result
\begin{equation}
\label{neweq2}
\EE C(w_{t+1}) - C(w^*) \leq
\end{equation}
\begin{equation*}
[\EE C(w_t) - C(w^*)] (1 - a_t m/D) + A K M^2 a_t^2.
\end{equation*}
We are now in a position to state our result:
\begin{theorem}
\label{thm:arate}
Let $C:\mathbb{R}^n\to\mathbb{R}$ be a twice differentiable cost function with
unique minimum $w^*$. Assume that
\vspace{-1ex}
\begin{enumerate}
\setlength{\itemsep}{0pt}
\item Conditions C.1--C.6 from Theorem~\ref{mainthm} hold with $B=0$ in C.4.
\item Equation~\eqref{cond} holds.
\item $a_t = \frac{\tau}{t}$ with $\tau>D/m$.
\end{enumerate}
Then $\EE C(w_t) - \inf_{\tilde{w}} C(\tilde{w})$
is equivalent to $\frac{1}{t}$ as $t\to\infty$.
\end{theorem}
\begin{proof}
\citet[A.4]{BotLec05} state: if $u_t=[1 - \frac{\alpha}{t} + o(1/t)]
u_{t-1} + \frac{\beta}{t^2} + o(1/t^2)$ with $\alpha>1$ and
$\beta>0$, then $tu_t\to\frac{\beta}{\alpha-1}$.
Theorem~\ref{thm:arate} follows from \eqref{neweq2} by setting
$u_t=\EE C(w_t) - C(w^*)$.
\end{proof}

\section{Quasi-Newton Methods}
\label{sec:bfgs}

Quasi-Newton methods are optimization methods with updates of the form
$ w_{t+1} \leftarrow w_t - a_t \,B_t \nabla_{w} C(w_t)$,
where $a_t > 0$ is a scalar gain (typically set by a line search)
and $B_t$ a positive-semidefinite scaling matrix. If $B_t =
 I$, we have simple gradient descent; setting
$B_{t}$ to the inverse Hessian of $C(w)$ we recover Newton's
method. Inverting the $k\times k$ Hessian is a
computationally challenging task if $k$ is large; quasi-Newton
methods reduce the computational cost by incrementally maintaining
a symmetric positive-definite estimate $B_t$ of the inverse Hessian
of the objective function.

\subsection{(L)BFGS}

The BFGS algorithm \citep{NocWri99} was developed
independently by Broyden, Fletcher, Goldfarb, and Shanno in 1970.
It incrementally maintains its estimate $B_t$ of the inverse Hessian
via a rank-two update that minimizes a weighted Frobenius norm
$\|B_{t+1} - B_t\|_W$ subject to the \emph{secant equation}
$s_t = B_{t+1} y_t$, where $s_t:= w_{t+1} - w_{t}$ and
$y_t := \nabla_{w} C(w_{t+1}) - \nabla_{w}C(w_{t})$ denote the
most recent step along the optimization trajectory in parameter and
gradient space, respectively. LBFGS is a limited-memory (matrix-free)
version of BFGS.

\subsection{Online (L)BFGS}

Recently developed online variants of BFGS and LBFGS, called
oBFGS \emph{resp.}\ oLBFGS, are amenable
to stochastic approximation of the gradient \citep{SchYuGue07}.
The key differences between the online and batch algorithms
can be summarized as follows:

The gradient of the objective is estimated from small
samples (mini-batches) of data. The difference $y_t$ of
gradients is computed with gradients for the same data sample, \ie for the same function $C(\cdot,z_t)$.
Line search is replaced by a gain sequence $a_t$.
A \emph{trust region} parameter $\lambda$ is introduced  modifying the algorithm to estimate $(H_t + \lambda I)^{-1}$, where $H_t$ is the Hessian at iteration $t$; this prevents the largest eigenvalue of $B_t$ from
exceeding $\lambda^{-1}$.

See \citet{SchYuGue07} for a more detailed description of the oLBFGS and oBFGS algorithms,
and for experimental results on quadratic bowl objectives and conditional random fields (CRFs), which
are instances of risk minimization.
\begin{remark}
To guarantee a uniform lower eigenvalue bound for the updates of \citet{SchYuGue07}
we would have to use $B_t + \gamma I$ for some $\gamma > 0$, effectively interpolating between
o(L)BFGS as defined by \citet{SchYuGue07} and simple online gradient descent. This lower bound
is not needed for convergence \emph{per se} but to prove that the convergence is to the
minimum.
\end{remark}

\subsection{Filtering}
\citet[Chapter 14]{KaiSayHas00} present assumptions in control theory that imply either
convergence of the matrices $B_t$ or upper and lower eigenvalue bounds. The relevant control
theory is too extensive and complicated to be reviewed here and we therefore only point out the connection.

\section{Expert Advice with Bregman Divergences}\label{sec:expert}
We now compare the updates in the expert advice framework by \citet{AzoWar01} to
the SGD updates that would result from a non-linear reparametrization.

As outlined in section \ref{breg} the difference between the update
\begin{equation}\label{theBregupdate}
w_{t+1}=h^{-1}(h(w_t)-a_t\nabla L_t(w_t)).
\end{equation}
and performing stochastic gradient descent with respect to a new variable $\gamma=h(w)$
is that the latter would include matrices $B_t$ which are the inverse Jacobians of $h$ at $w_t$, i.e. the inverse
Hessian of $H$ where $\nabla H=h$. The update $\gamma_{t+1}=\gamma_t-a_t\nabla_\gamma \tilde{L}_t(\gamma_t)$ where $\tilde{L}_t(\gamma)=L_t(h^{-1}(\gamma))$, expressed with respect to
$w$ looks as follows
\begin{equation}\label{theBregupdate2}
w_{t+1}=h^{-1}(h(w_t)-a_tB_t\nabla L_t(w_t)).
\end{equation}
Therefore \eqref{theBregupdate} can be written as
\begin{equation}
\gamma_{t+1}=\gamma_t-a_tB_t^{-1}\nabla_\gamma \tilde{L}_t(\gamma_t).
\end{equation}
The main point of the reparametrization would be to change variables so that SGD converges with an
optimal rate for the new objectives. We would like the new objective to be approximately quadratic.
Assuming that we have chosen the transfer function in such a manner that our main theorems apply
to the new objective function, it remains to check the scaling matrix condition.

To satisfy the conditions for the scaling matrices we need the Jacobian of the transfer function to
have upper and lower eigenvalue bounds. In the literature that these methods are studied in, an assumption of uniformly bounded gradients $\nabla L_t(w_t)$ \citep{AzoWar01} is often used, or the parameters are
restricted to a compact set \citep{Zinkevich03}.
In that case the conditions on scaling matrices become easier to satisfy:
in any compact set, popular functions like $e^\theta$,  $(1+e^\theta)^{-1}$, or other sigmoid functions
(though not, \emph{e.g.}, $\theta^3$ which is flat at the origin) have derivatives that are bounded from above and below by a strictly positive constant. 

\section{Conclusion}
\label{sec:conclusion}

We provide a variable metric stochastic approximation theory which implies convergence for
stochastic gradient descent even when the gradients are calculated with respect to variable metrics.
Metrics are sometimes changed for optimization purposes, as in the case of using quasi-Newton
methods. Our main theorems imply convergence results for online versions of the BFGS and LBFGS optimization  methods. Kalman filters are a class of well-known algorithms that can be viewed as an
online Newton method for the special case of square losses, since the procedure at every step performs
a gradient step where the gradient is defined using the Hessian of the loss for the examples seen so far.
Finally we investigate the task of learning from
expert advice where Bregman divergences are frequently used to achieve updates that are suitable for the
task at hand. We interpret the resulting updates as stochastic gradient descent in a space that has undergone
a non-linear reparametrization, and where we use different metrics depending on the point we are at.


\subsubsection*{Acknowledgements}

The authors are very grateful to Leon Bottou at NEC Labs in Princeton, NJ for his help with the main theorem.

The first author is funded by NICTA which is funded through the Australian Government's \emph{Backing Australia's Ability initiative}, in part through the Australian Research Council.

\bibliographystyle{../abbunnat}
\bibliography{../../../bibfile/bibfile}
\end{document}